\documentclass{article}
\usepackage[a4paper, margin=1.5cm]{geometry}
\usepackage{graphicx}
\usepackage{setspace}
\usepackage{caption}
\usepackage{booktabs}
\usepackage{multicol}
\usepackage{mathrsfs}
\usepackage{amsmath}
\usepackage{float}       
\usepackage{stfloats}    
\usepackage{cuted}       
\usepackage{lipsum}      
\usepackage{authblk}
\usepackage[mathscr]{euscript}
\usepackage{xfp}
\usepackage{cite}
\usepackage{xcolor,soul}
\newcommand\SupplementaryMaterials{%
  \xdef\presupfigures{\arabic{figure}}
  \xdef\presupsections{\arabic{section}}
  \renewcommand\thefigure{S\fpeval{\arabic{figure}-\presupfigures}}
  \renewcommand\thesection{S\fpeval{\arabic{section}-\presupsections}}
}

\title{\textbf{Synchronous polarization switching at sub-coercive fields through stochastic resonance in ferroelectric thin-film capacitors}}

\author[1$^\#$]{Vivek Dey}
\author[2$^\#$]{Thejas Basavarajappa}
\author[3]{Manikantan R.S.}
\author[1,3]{Kevin Renji Jacob}
\author[1,3]{Jonnalagadda Nikhila}
\author[3]{Arvind Ajoy}
\author[1]{Pavan Nukala}

\affil[1]{Centre for Nanoscience and Engineering, Indian Institute of Science, 560012, India}
\affil[2]{School of Electrical and Computer Engineering, Cornell University, Ithaca, NY 14853, USA}
\affil[3]{Department of Electrical Engineering, Indian Institute of Technology Palakkad, 678623, India
\vspace{1em}}
\affil[$\#$]{These authors contributed equally to this work}
\affil[ ]{Corresponding authors: pnukala@iisc.ac.in, arvindajoy@iitpkd.ac.in}

\date{}                     
\date{}

\begin{document}
\onehalfspacing

\maketitle

\begin{abstract}
\noindent
Stochastic resonance (SR) is a phenomenon by which the presence of noise in a non-linear system allows for detection of a weak sub-threshold signal, or in a bi-stable system allows for sub-coercive switching between the two states. Simple theory suggests that SR occurs when the Kramers rate ($r_{k}$) of the bistable system, which is a function of noise and applied voltage, is twice the drive frequency ($f_{signal}$). Here, we demonstrate the synchronous  switching of polarization with a sub-coercive voltage waveform, in a thin film ferroelectric lead zirconium titanate (PZT) capacitor through SR. We employ independent figures of merit (FOM) such as cross-covariance, output power and signal-to-noise ratio to experimentally identify the optimal noise for synchronous switching. We further experimentally measure the Kramers time in the ferroelectric, and show that FOMs indeed peak near the noise predicted by the SR condition. We also model the device characteristics using the stochastic Time Dependent Landau Ginzburg (TDGL) formulation, and capture the experimentally observed polarization switching  under application of sub-coercive voltage, assisted by noise. Finally, we show a proof-of-concept implementation of detecting sub-threshold frequency-shift-key signals (FSK) in noisy communication channels using our ferroelectric PZT devices.

\end{abstract}
\vspace{1cm}

\section{Introduction}
Noise, in general, has a detrimental effect on system performance as it introduces distortions to the signal under study. However, in a non-linear system, noise can counterintuitively facilitate the retrieval of weak, sub-threshold signals. This phenomenon is referred to as  stochastic resonance (SR). Some of the first studies of SR and its mechanisms, were employed to explain long term climatic transitions and periodic recurrence of Earth’s ice ages \cite{JournalofPhysicsAMathematicalandTheoretical1981, Tellus1982A, Tellus1982B, JournalofStatisticalPhysics1993}.  Since then, SR has been demonstrated/engineered in various systems, ranging from biological to physical systems
\cite{Nature1993,  Nature1996, Nature1999, Journaloftheoreticalbiology2002, PhysicalReviewE1994, PhysicalReviewLetters1988, Naturecommunications2020, AppliedPhysicsLetters1995, ChaosSolitonsFractals2021,NPJ_2024,JapaneseJournalofAppliedPhysics2011, JournalofAppliedPhysics1995,   
PhysicsLettersA1983, PhysicalReviewLetters1991 }. For example, in sensory neurobiology, SR has been used to explain animal behaviors contributing to evolutionary success, such as the paddlefish, which uses optimal noise to efficiently detect its food (the zooplankton Daphnia), a task that would otherwise be impossible due to limited vision caused by poor light and turbid water \cite{Nature1999, Journaloftheoreticalbiology2002}. 

\begin{figure*}[t]
\centering
\includegraphics[scale = 1]{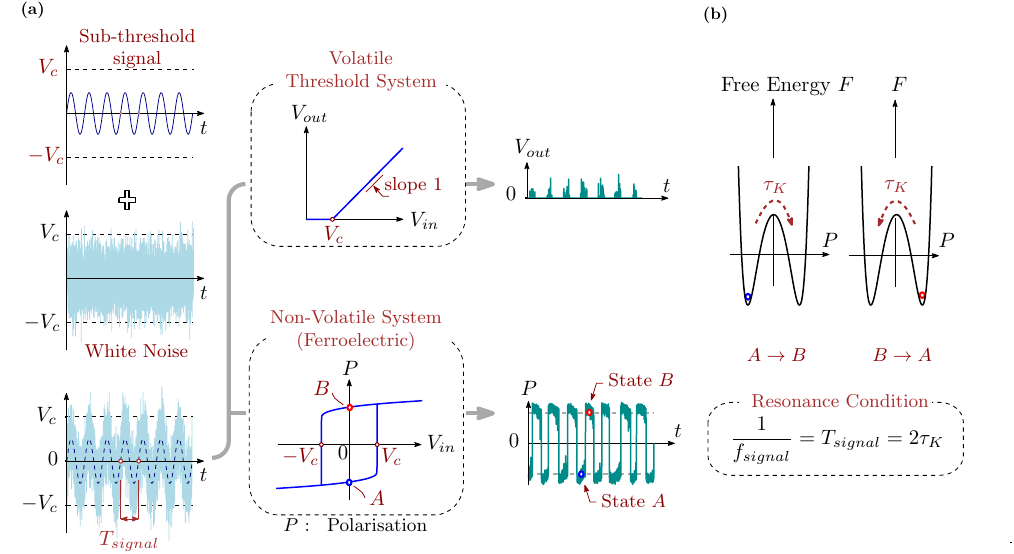}
\caption{\textbf{Concept of Stochastic Resonance.} (a) Illustration of  stochastic resonance in  volatile/threshold and non-volatile bistable (e.g. ferroelectric) systems. Addition of white noise helps the system to cross the threshold, enabling the detection of sub-threshold signals. (b) Resonance condition in a ferroeletric:  state of the system switches twice ($2 \times \tau_K$) within one period of signal ($T_{signal}$).  This relationship helps to determine the optimum noise needed for SR.}
\label{fig:kramersproblem}
\end{figure*}

Non-linear electronic devices/systems  with the potential to demonstrate SR can be broadly classified into two types -- volatile (i.e., systems with a  single threshold) and non-volatile (i.e., bi-stable systems with two distinct thresholds for change of state from $A \rightarrow B$ and $B \rightarrow A$), as illustrated in Fig. \ref{fig:kramersproblem}. In volatile systems, SR has been implemented in the context of detecting and amplifying subthreshold signals -- for example, tunnel diodes \cite{PhysicalReviewE1994}, ring lasers \cite{PhysicalReviewLetters1988}, photodetectors \cite{Naturecommunications2020} and optical systems \cite{AppliedPhysicsLetters1995}, memristors \cite{ChaosSolitonsFractals2021, NPJ_2024}, carbon nanotube transistors \cite{JapaneseJournalofAppliedPhysics2011}, superconducting Josephson junctions \cite{JournalofAppliedPhysics1995}. A few demonstrations of SR in non-volatile systems have also been reported for e.g., electronic circuits such as a Schmitt trigger \cite{PhysicsLettersA1983}, in a bi-stable electro-paramagnetic-resonance system \cite{PhysicalReviewLetters1991} and in bulk Triglycine Sulfate (TGS) single crystals \cite{Ferroelectrics2000}. Here, an \emph{optimum} amount of noise causes the output to switch synchronously with the sub-threshold signal.

Ferroelectrics represent a prototypical non-volatile system described by a double-well energy landscape, as shown in Fig. \ref{fig:kramersproblem}(b). In these systems, at sub-coercive (i.e sub-threshold) electric fields, noise dictates the rate \cite{Physica_1940} at which the system jumps from one well to another (called  Kramers rate $r_K$, which is the inverse of Karmers time $\tau_K$ ). As illustrated by Fig. \ref{fig:kramersproblem}(b), $T_{signal} = 2\tau_K$ (where $T_{signal}$ is the period of the weak signal) defines the condition for SR. Kramers time is a function of input noise -- hence, SR occurs when an optimal amount of noise is added, causing the  polarization to switch synchronously even at sub-coercive fields. Thus far,  SR in ferroelectrics has  only been measured in bulk Triglycine Sulfate (TGS) single crystals \cite{Ferroelectrics2000}. However, weak ferroelectricity in this system makes it difficult to distinguish the dielectric hysteresis loss from the ferroelectric polarization switching. 

In this study, we demonstrate SR (sub-coercive field synchronous polarization switching) in  robust, thin film ferroelectric Lead Zirconate Titanate (PZT), with comprehensive experimental results supported by theory and simulation. We perform SR experiments in two different representative PZT ferroelectric capacitors (with coercive voltages $V_{cA}$ and $V_{cB}$ respectively), with the input amplitude $V_{signal} = 0.75 \ V_{cA}$ (for Sample-A) and $V_{signal} = 0.6 \ V_{cB}$ (for Sample-B). We identify three figures of merit for sub-coercive field polarization switching i.e., output power, cross-covariance and signal-to-noise ratio, all of which peak at an optimal amount of noise. We independently measure Kramers time  and show that this optimal noise indeed corresponds to SR. Our detailed numerical simulations using stochastic time-dependent Landau-Ginzburg Devonshire (TDGL) theory \cite{Landau1937, Ginzburg1945, Devonshire1949} successfully reproduce the experimentally observed polarization switching  under application of  sub-coercive fields aided by noise. A thin film ferroelectric platform is very conducive for device applications. We present a proof-of-concept implementation of sub-threshold frequency-shift-key (FSK) signal recovery using SR in our devices. This highlights the potential of our approach to improve signal detection in noisy communication channels, such as underwater communication.


\begin{figure*}[t]
\centering
\includegraphics[width=\textwidth]{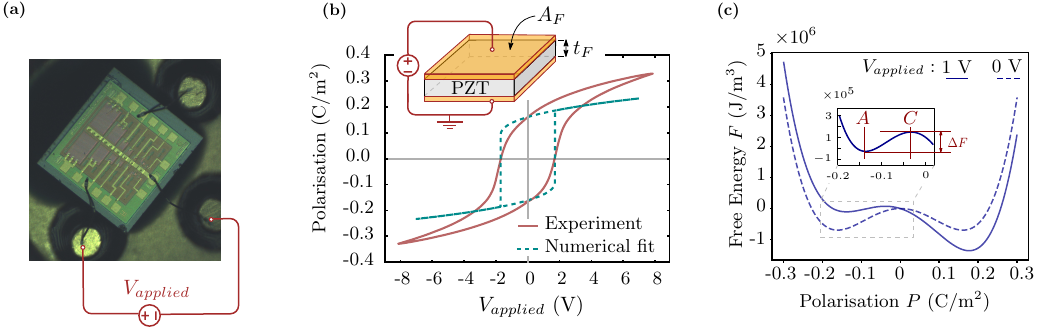}
\caption{\textbf{PZT device characterization} (a) Optical image of our representative PZT sample with connected probes for SR measurement. (b) Polarization vs. voltage hysteresis loop measured with a $1$ Hz bipolar signal for the PZT device (solid brown), and the fit using a single domain (indicated in cyan dash). (c) Estimated double well from Landau theory. Inset shows the double well at different voltage biases, including an estimate of the barrier height $\Delta F$.}

\label{fig:PEloop2}
\end{figure*}

\begin{table}[b]
\centering
{%
\begin{tabular}{lcc}
\hline
\textbf{Parameter} & \textbf{Sample - A} & \textbf{Sample - B} \\
\hline
Thickness, $t_F$ (nm) & $255$ & $255$ \\
Area, $A_F$ ($\mu$m$^2$) & $4\times10^3$ & $10^5$ \\
Coercive Voltage, $V_c$ (V)& $2.1$ & $1.7$ \\
Remanent Polarization, $P_r$ (C/m$^2$) & $0.17$ & $0.161$ \\
Landau Coefficient, $\alpha$ (mF$^{-1}$) & $-6.29 \times 10^7$ & $-5.38 \times 10^7$ \\
Landau Coefficient, $\beta$ (m$^5$F$^{-1}$C$^{-1}$) & $1.09 \times 10^9$ & $1.04 \times 10^9$ \\
Resistivity, $\rho$ ($\Omega \cdot$ m) & $1367$ & $390$ \\
\hline
\end{tabular}
}
\caption{Material Parameters for Ferroelectric Samples}
\label{tab:TableParam}
\end{table}

\section{Electrical Characterization and Landau-Based Modeling of PZT Thin-Film Capacitors}
\label{sec:parameterExtraction}
Table \ref{tab:TableParam} lists the geometry  (Fig. \ref{fig:PEloop2}(a)) and material parameters for both the PZT (PbZr$_{0.2}$Ti$_{0.8}$O$_3$)  thin film capacitors (Sample-A and Sample-B). Fig. \ref{fig:PEloop2}(b) shows the representative polarization vs. applied voltage hysteresis obtained from the Sample-B (see Fig. \ref{fig:PEsampleA} for Sample-A), which is measured using a 1 Hz standard bipolar signal.

Next we model the ferroelectric bistable potential well of PZT  using a single domain Landau theory. 
The energy landscape of the ferroelectric can then be described  using Landau free energy density $F$ (J/m$^3$) as
\begin{align}
F = \alpha {P} ^ {2} + \beta {P} ^ {4} - P  E
\label{eq:ferropotential}
\end{align}
where $P$ is the ferroelectric polarization (C/m$^2$), $E$ is the applied electric field and $\alpha = \dfrac{-3\sqrt{3}E_c}{4P_r}$, $\beta = \dfrac{3\sqrt{3}E_c}{8P_r^3}$  \cite{IEEEtransactionsonelectrondevices2016} are Landau coefficients. From these parameters we extract a symmetric double well potential for the two samples A and B with a barrier height of $\Delta F = 9 \times 10^5$ J/m$^{3}$ and $6.89 \times 10^5$ J/m$^{3}$ respectively (see Fig. \ref{fig:PEloop2}(c) and Fig. S1).
The PE loop obtained by solving the time dependent Ginzburg-Landau equation (TDGL)(eq. 2a) is shown in Fig. 2b (cyan colored dashed line). Note that the mismatch between the experimentally obtained and simulated PE loop is a result of not considering the multidomain switching. However, single domain switching model is good enough to quantitatively describe the SR, as we discuss later.

The polarization response of the PZT capacitor when excited by a deterministic signal $V_{signal}(t)$ and a noise source $V_{noise}(t)$, with root-mean-square value $V_{noise\; rms} = \sqrt{\langle V^2_{noise}(t) \rangle}$ can be obtained by solving the stochastic time dependent Ginzburg-Landau equation (TDGL) \cite{arXiv2021}, 
\begin{subequations}
\begin{align}
\rho \dfrac{\partial P}{\partial t}  & = - \dfrac{\partial F}{\partial P} + \xi (t) \\
\text{ with }\xi (t) &= \left( \sqrt {2 \rho D_{int}} + \sqrt {2 \rho D_{ext}}\right)  \dfrac{dW(t)}{dt} \\
\text{ where  } D_{int} & = \dfrac{k_BT}{t_F A_F}, \\
  D_{ext} & = \dfrac{V^2_{noise\; rms}}{2 R_F} \dfrac{1}{\Delta f_{BW}} \dfrac{1}{t_F A_F} \\
  \text{ and } R_F & = \dfrac{\rho t_F}{A_F}
\end{align}
\label{eq:TDGL}
\end{subequations}
\noindent Here, $t_F$ is thickness of the PZT capacitor, $A_F$ is the area, and $\rho$ ($\Omega$m) is the switching resistivity, a dampling/dissipative term, estimated from Kramers time. The stochastic nature of the input is described by $\xi(t)$, while field $E(t)$ captures the deterministic part of the input. $\xi(t)$ is related to an  underlying Brownian motion or Weiner process $W(t)$, with $D$ being the diffusion coefficient. Specifically, $D_{int}$ describes  fluctuations due to internal thermal noise ($k_B$ is Boltzmann's constant and $T$ is temperature), while $D_{ext}$ captures the effect of the external noise. For our case,  $D_{int} \ll D_{ext}   $ at room temperature in the samples that we measure, and hence can be ignored. A more rigorous formulation can be found in \cite{arXiv2021, PhysicalReviewB1994, JournalofPhysicsMathematicalandTheoretical2020, PhysicalReviewApplied2017, PhysicalReviewB2016, risken1996fokker} (also detailed explanation in Supplementary Note 1). 

The TDGL equation eq. (\ref{eq:TDGL}) is analogous to the classic particle escape problem under the strong damping limit \cite{Physica_1940, ChemicalPhysicsLetters2000}. The rate of escape of the polarization state (from a well $A$) over the energy barrier $\Delta F$ at $C$ (as shown in the inset in Fig. \ref{fig:kramersproblem}(b)), is known as the Kramers rate $r_K$ and given by 
\begin{align}
    r_K = \frac{1}{\tau_K} = \frac{\sqrt{|F''(P_A) F''(P_C)|}}{2 \pi \rho} \exp{\left(-\dfrac{\Delta F}{ D_{ext}} \right)}
    \label{eq:KramersTime}
\end{align}
where we use the approximation $D_{int} \ll D_{ext}$. Kramers time $\tau_K$ represents the average \emph{first passage time} for the system over the barrier. From eq. (\ref{eq:KramersTime}), note that $D_{ext} / \Delta F$ hence provides a good measure to quantify the amount of external noise being added to the system \cite{arXiv2021}.  As illustrated in Fig. \ref{fig:kramersproblem}(b), optimum noise for SR corresponds to the situation when the state of the system switches twice ($2 \times \tau_K$) within one period of signal ($T_{signal}$). 

To solve eq. (\ref{eq:TDGL}) numerically, we use the  discretized form as 
\begin{align}
P^{[i]} = P^{[i-1]} - \frac{\Delta t}{\rho} \left[\frac{\partial F}{\partial P}\right]^{[i-1]} + \sqrt{\frac{2D_{ext}}{\rho}} \cdot \left[\Delta W\right]^{[i-1]}
\label{eq:TDGL_Discrete}
\end{align}
where $\Delta t$ $(10$ ns) is the simulation timestep, $[i]$ represents the $i^{th}$ time. Following the definition of the Weiner  process, $\Delta W$, representing the increment in the random noise at every simulation timestep, is extracted from a normal distribution with mean $0$ and variance $\Delta t$. We solve eq. (\ref{eq:TDGL_Discrete}) using the Euler-Maruyama method \cite{ActaNumerica1999} to obtain the time dependent polarization under the application of weak sub-threshold signal and external noise.  An appropriate filter is introduced to ensure that the bandwidth of the noise is limited to the experimental bandwidth of the measurement setup $\Delta f_{BW}$.

\begin{figure*}[t]
\centering
\includegraphics[width=\textwidth]{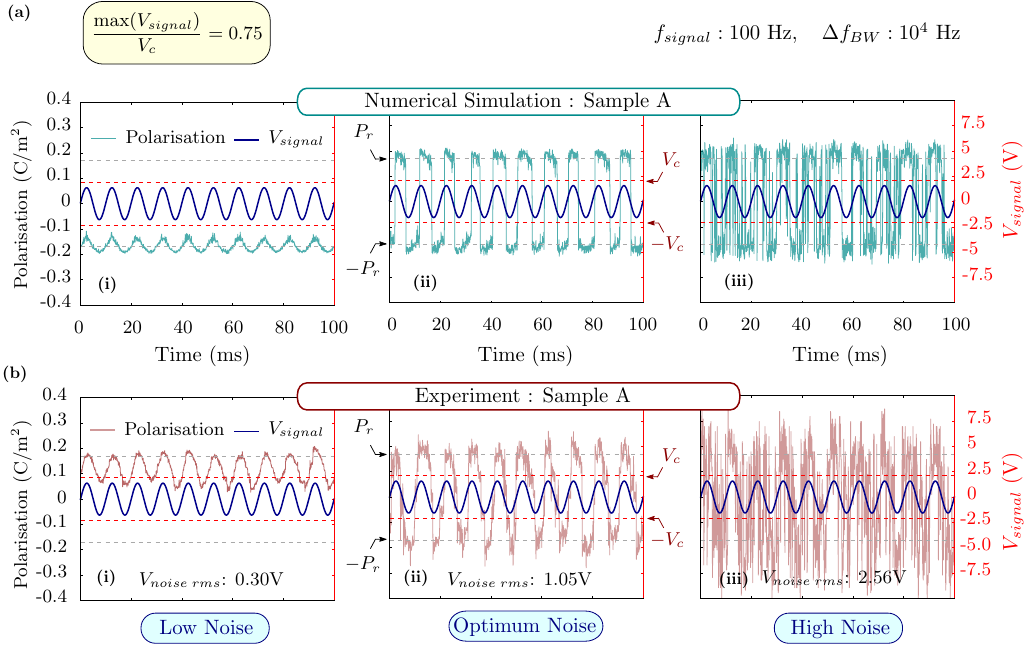}
\caption{\textbf{Simulation and experimental polarization switching response at different noise strength}. (a) Simulated and (b) Experimental polarization switching recorded from the PZT capacitor Sample A at different noise strength. At very low noise ($V_{noise\; rms} = 0.30$ V) the polarization modulates in one of the wells; at high noise ($V_{noise\; rms} = 2.56$ V) the polarization switches randomly between the two wells. However, at an optimum noise in between, $V_{noise\; rms} = 1.05$ V, the polarization switches quasi-periodically and is synchronous with the input signal.}
\label{fig:PtsampleA}
\end{figure*}

Fig. \ref{fig:PtsampleA}(a)  shows the simulated polarization vs. time plots for Sample-A at different noise strengths, $V_{noise\; rms}$, for $100\,\text{Hz}$ signal at sub-coercive voltage of 0.75 $V_{cA}$ (see Fig. S3 and S4 for $75\,\text{Hz}$ and $150\,\text{Hz}$, and Fig. S5 for Sample-B at $100\,\text{Hz}$). For the $100 \,\text{Hz}$ signal, we note that at a low $V_{noise\; rms} = 0.30$ V the polarization does not switch, and at a very high $V_{noise\; rms} = 2.56$ V, the polarization switches randomly and asynchronously with the drive voltage signal. However, at $V_{noise\; rms} = 1.05$ V, a value in between, the polarization switches quasi-periodically in response to the input sub-coercive voltage, indicating stochastic resonance.

To rigorously characterize the optimal noise for synchronous polarization switching, we define the following figures of merits:
\begin{enumerate}
\item \textbf{Cross-covariance (cov)}: Cross covariance between the reference polarization signal $P_1(t)$, obtained with a super-coercive sinusoidal input voltage signal (at an amplitude $V_{signal} = 1.5 \ V_c$, without any noise), and the observed polarization signal $P_2(t)$ at sub-coercive voltages with noise is calculated as,
\begin{align}
\text{cov}(P_1, P_2) = \frac{1}{N-1} \sum^N_{i=1} (P_1 - \mu_1)\cdot(P_2 - \mu_2)
\label{eq_FOM1}
\end{align}

where N is the number of samples in the voltage signal and $\mu_1$ and $\mu_2$ are the average values of the two signals $P_1$ and $P_2$.

\item \textbf{Output power (OP)}: The total power in the  output polarization signal within a frequency bandwidth ($\Delta f_{signal}$, illustrated in Fig. \ref{fig:SampleA75Hz}(d)) around $f_{signal}$  is calculated as
\begin{align}
 \mathscr{P}_{signal} = \int \limits_{f_{signal}-\Delta f_{signal}/2}^{f_{signal}+\Delta f_{signal}/2} \!  \!  \!  \!   \!  \!   \!  \!  \!  \!  \!   \!  \!   \ \!  \!   \!  \!  \!   \! \text{PSD}\{P(f)\} df 
 \label{eq_FOM2}
\end{align}
with PSD$\{P(f)\}$ being the power spectral density of the output polarization. 

\item \textbf{Signal to noise ratio (SNR)}:  is obtained as the ratio of  power of the signal $\mathscr{P}_{signal}$ and noise power within selected a bandwidth  $\Delta f_{noise}$, as illustrated in Fig. \ref{fig:SampleA75Hz}(d) as,
\begin{subequations}
\begin{align}
\text{SNR} &=  \frac{\mathscr{P}_{signal}}{\mathscr{P}_{noise}} \text{ with}  \\
\mathscr{P}_{noise} &= \int\limits_{f_{signal}-\Delta f_{noise}}^{f_{signal}+\Delta f_{noise}}  \!  \!  \!  \!   \!  \!   \!  \!  \!  \!  \!   \!  \!   \!
\text{PSD}\{P(f)\} df \; - \; \mathscr{P}_{signal}
\end{align}
\label{eq_FOM3}
\end{subequations}

\end{enumerate}

\begin{figure*}[t]
\centering
\includegraphics[width=\textwidth]{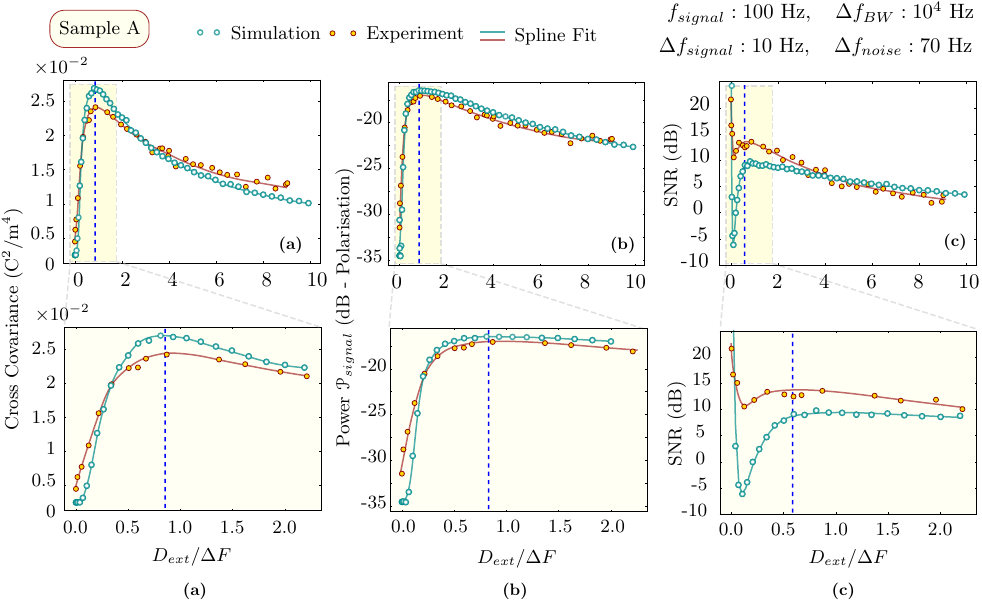}
\caption{\textbf{Figure of merits to quantify SR: Cross covariance, Signal Power, Signal to noise ratio (SNR)}. We compare the prediction from the numerical simulation results with the experimentally measured results for Sample A. (a) Cross-covariance, (b) output power, and (c) SNR for simulation (cyan) and experimental (brown) results. The blue dashed lines show the position of the peaks observed in the experimental data. The experimental and simulated results matches well peaking at $V_{noise\; rms} = 1.05$ V or $ D_{ext} / \Delta F = 0.87 \pm 0.21 $. The bottom panel shows a zoomed-in view of the yellow region highlighted in the top-panel figures.}

\label{fig:SRsampleA100}
\end{figure*}

Fig. \ref{fig:SRsampleA100} shows the three different metrics: cross-covariance, output power and snr (cyan color) calculated from the simulated polarization switching response of Sample A. We find that for the simulated potential well the three metrics peak at $V_{noise\; rms} = 1.05$ V or $ D_{ext} / \Delta F = 0.87 \pm 0.21 $ for Sample-A, which defines the optimal noise for sub-coercive field switching.
\section{Experimental demonstration of stochastic resonance}

Next, we apply sub-coercive voltage signals to the PZT capacitors in the presence of noise and record the polarization switching \( P(t) \). For this, we generate periodic voltage signals with amplitudes of \( 0.75\,V_{cA} \) for sample A at different frequencies: $75\,\text{Hz}$, $100\,\text{Hz}$, and  $150\,\text{Hz}$, (reported in Fig. \ref{fig:SampleA75Hz}, Fig. \ref{fig:SRsampleA100} and  Fig. \ref{fig:SampleA150Hz} respectively) and amplitude of 0.6 $V_{cB}$ at 100 Hz for Sample-B. We add white Gaussian noise with a bandwidth \( \Delta f_{\text{BW}} \) and vary the RMS noise amplitude $V_{noise\; rms}$ before feeding the signal into the devices (more details in Fig. \ref{fig:Noise_gen} and Supplementary Note 2). The devices were poled before every measurement cycle to ensure that the starting state is the same well in the double-well potential (for poling protocols, see Supplementary Note 3). Fig. \ref{fig:PtsampleA}(b) shows the polarization response of Sample-A for the 100 Hz signal  at different $V_{noise\; rms}$. The remnant polarization  $P_r$ and $-P_r$ obtained from the polarization loop (Fig. \ref{fig:PEsampleA}(a)) are shown as gray dotted lines in Fig. \ref{fig:PtsampleA}, and any polarization switching event will involve a change in polarization $P < -P_r$ to $P > P_r$ or vice versa. Changes in polarization around one of these limits are just the result of dielectric charging and not polarization switching. Using these identification markers, we clearly see that quasi-periodic switching occurs at an optimal value of noise, $V_{noise\; rms} = 1.05$ V (Fig. \ref{fig:PtsampleA}(b)(ii)), with large noise $V_{noise\; rms} = 2.56 $ V (Fig. \ref{fig:PtsampleA}(b)(iii)) asynchronously switching the device, and small noise $V_{noise\; rms} = 0.30$ V just leading to dielectric charging (Fig. \ref{fig:PtsampleA}(b)(i)). 
We further estimate all the independent figures of merit (Fig. \ref{fig:SRsampleA100}(a)-(c)) which all peak at an optimal value of noise, that exactly matches with the optimal noise predictions from our stochastic TDGL simulations (SR noise: $V_{noise\; rms} = 1.05 \ V$ or $ D_{ext} / \Delta F = 0.87 \pm 0.21 $). Such a remarkable agreement between the model and experiment emphasizes the robustness of our phenomenological modeling framework. Similar results are also obtained with Sample-B, as shown in Fig. \ref{fig:SRsampleB}.

\begin{figure*}[!t]
\centering
\includegraphics[width=\textwidth]{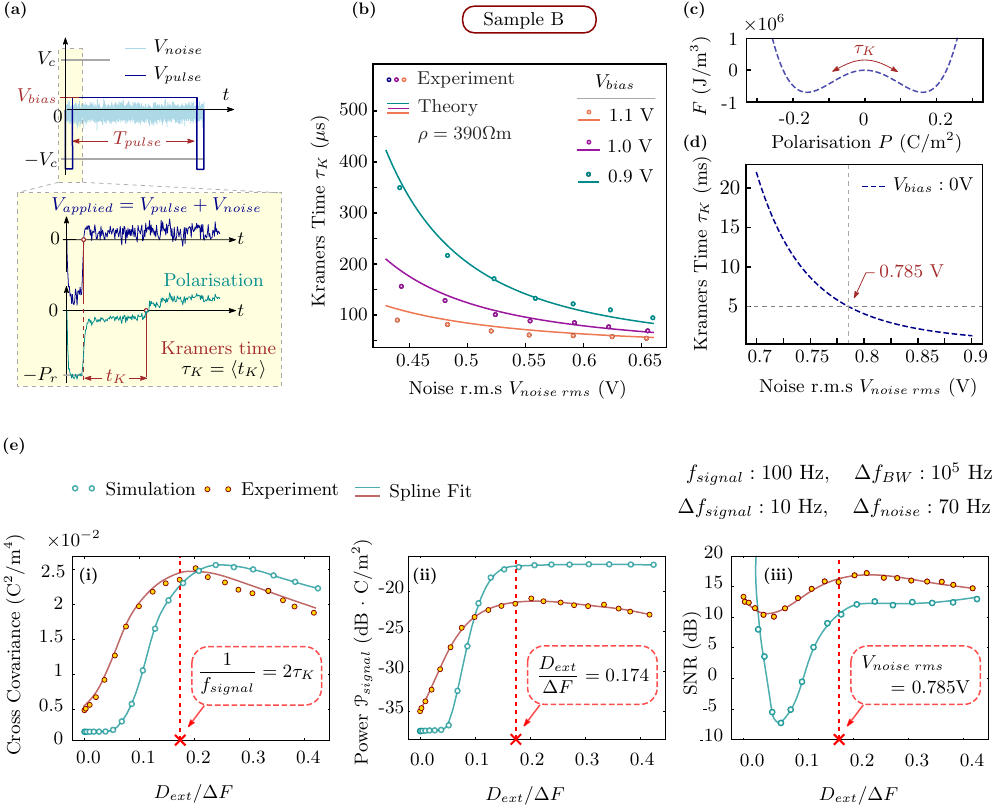}
\caption{\textbf{Kramers time.} (a) Experimental protocol to measure Kramers time of the PZT sample. The capacitor is \emph{reset} to the negative polarization state $-P_r$. The \emph{first passage time}, $t_k$, is measured from the time of application of the sub-coercive \emph{set} pulse, to the time when polarization crosses zero. Averaging over multiple \emph{reset-set} pulses yields Kramers time $\tau_K$. (b) Measured Kramers time (dots) for different amplitudes of noise, along with the theoretical fit (solid) obtained with  $\rho$ value of $390$ $\Omega$m. (c) The computed double-well without bias. (d) The theoretical estimate for Kramers time in a symmetric double well. (e) Simulated and experimental figure of merits: (i) cross-covariance, (ii) output power and (iii) SNR. The red vertical dashed lines show the optimum noise (represented in terms of $D_{ext}/\Delta F$ and $V_{noise \;rms}$) for SR from the estimate of Kramers time.}
\label{fig:kramerstime}
\end{figure*}

\section{Estimation of Kramers time from double well of the PZT capacitor}
To answer the question whether the optimal noise at which the FOMs peak corresponds to prediction from the intuitive SR condition (or not), we measured the Kramers time of our devices. Here, we present our representative measurements from Sample-B. Kramers time defines the timescale for switching of a double well system from one state to another, as discussed in Fig. \ref{fig:kramersproblem}(b). 
The strategy for measuring Kramers time is as follows: (a) We first measure the Kramers time in a biased double-well at various values of bias. The bias ensures that the transition $A \rightarrow B$ is significantly more likely than $B \rightarrow A$. (b) Then, we estimate the Kramers time  for the unbiased double well by using eq. (\ref{eq:KramersTime}), substituting the estimated barrier height $\Delta F$ from eq. (\ref{eq:ferropotential}). 
To this end, the device polarization is first \emph{reset} to state A (negative polarization $-P_r$) with a short-duration ($250$ $\mu$s), high negative voltage poling pulse. This is followed by  a \emph{set pulse} of a sub-coercive bias voltage for a duration of $T_{pulse} = 1750\ \mu$s.  We obtain the \emph{first passage time}, $t_K$, as the time from the start of the bias pulse to the time when polarization crosses zero. Given the stochastic nature of the process of switching, we perform 300 different such measurements (for each value of $V_{bias}$ and $V_{rms \; noise}$) and estimate the Kramers time $\tau_K = \langle t_K \rangle$ as the ensemble average over the 300 \emph{reset-set} pulses. Fig. \ref{fig:kramerstime}(b) presents the experimentally measured Kramers time (in dots) at different noise voltages compared with theoretical predictions (solid lines) from eq. (\ref{eq:KramersTime}) using  resistivity $\rho = 390 \ \Omega$m, which is the only fitting parameter.
Using this $\rho$ we estimate Kramers time vs noise for an unbiased double well potential (shown in Fig. \ref{fig:kramerstime}(c)). For an input signal frequency of 100 Hz ($T_{signal} = 10$ ms) SR is predicted to occurs at Kramers time of 5 ms. At $\tau_k = 5$ ms, $V_{noise\; rms} = 0.785 \ V$ from Fig. 5d.
The three figures of merit  from the experimental results of Sample-B at 100 Hz are shown in Fig. \ref{fig:kramerstime}(e) and peak at $V_{noise\;rms}$ = 0.78 V or $ D_{ext}/\Delta F= 0.171 \pm 0.009 $. These values are consistent with the predictions from the SR condition (shown as red vertical lines). Simulations using the TDGL are also well aligned with the experimental measurements. Please note that this is the first time report of direct measurement of SR through  Kramers time in a bistable ferroelectric system. Furthermore, we prove that our FOMs are robust independent measures of identifying SR.

\section{Detection of weak frequency shift key signals using SR}
\begin{figure*}[t]
\centering
\includegraphics[scale=0.9]{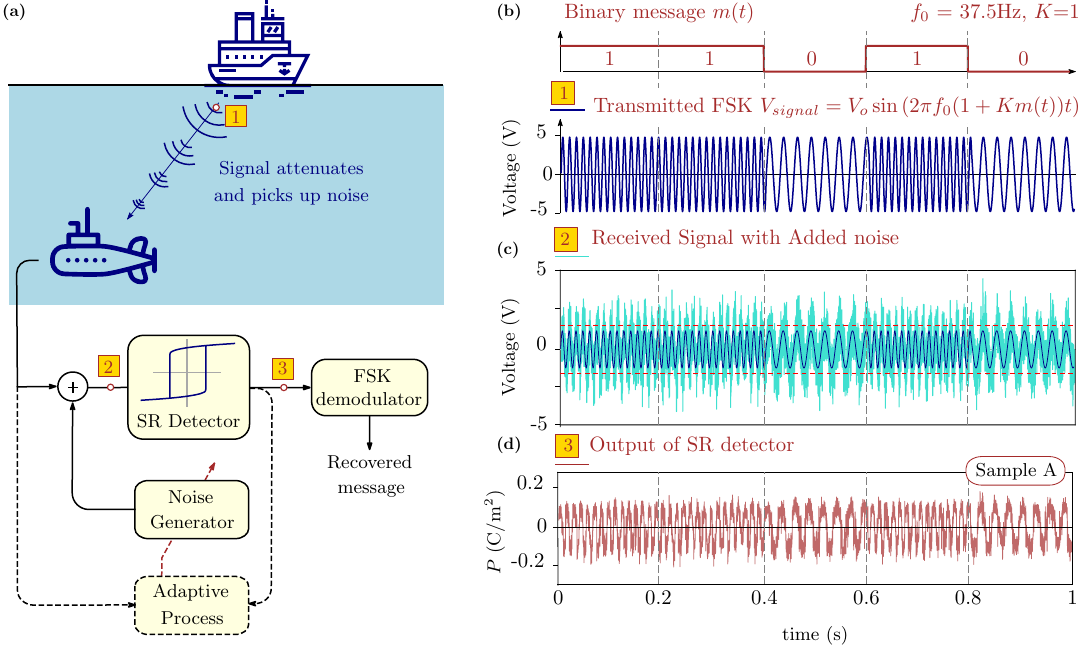}
\caption{\textbf{SR detector for weak signal recovery}. (a) Proof-of-concept SR based signal receiver for FSK reception in a noisy environment. Note that we do not implement the adaptive process in this work (b) The message signal $m(t)$ is encoded via the instantaneous frequency of the transmitted FSK signal, shown in (1); (2) The weak (sub-coercive) received signal along with added noise, (3) The polarization output of the SR detector in response to the noisy weak FSK signal (2).}
\label{fig:fsk}
\end{figure*}
Finally, we present a proof-of-concept demonstration of recovering sub-coercive signals carrying information using Frequency Shift Keying (FSK) modulation. This highlights the potential of our approach for enhancing signal detection in noisy communication channels, such as underwater communication. FSK is a method of digital signal modulation where binary data $m(t)$ is encoded in the frequency $f_{signal}(t)$ of a carrier signal and transmitted across a communication channel. FSK is widely used in applications such as radio communication, modems, and other wireless communication systems \cite{DUAN2005401}. A schematic for detection of weak FSK signal is shown in  Fig. \ref{fig:fsk}(a). In our demonstration, the FSK signal from the emitter consists of a burst of $37.5$ Hz and $75$ Hz sinusoids corresponding to bits $0$ and $1$ respectively (Fig. 6(b)). These signals attenuate in the channel and pick up noise, such that the effective signal amplitude comes below the threshold of the detector (as illustrated in Fig. 6(c)). We conceptualize our thin-film ferroelectric PZT device acting as an SR detector, to which the noisy, sub-coercive signal and additional noise from the noise generator are fed. The effective noise in the input to the SR detector hence includes both the channel noise and the added noise. At an optimal value of the total noise, our detector harnesses SR to generate a polarization signal with exactly the same instantaneous frequency as the transmitted FSK signal (Fig. \ref{fig:fsk}(b)), shown in Fig. 6(d), thus effectively recovering the sub-coercive FSK signal.

\section{Conclusions}
In this study, we demonstrate stochastic resonance (SR) in a bistable ferroelectric potential well by modulating the well with a weak sub-threshold periodic signal in the presence of white noise. We show that at an optimal noise strength, the polarization state of the PZT capacitors switches quasi-periodically in synchronization with the input signal frequency. We model the bistable potential well of PZT using Landau theory and simulate the polarization dynamics numerically solving the stochastic time-dependent Ginzburg-Landau (TDGL) equation using the Euler-Maruyama method. Our single-domain model shows excellent agreement with experimental results, accurately predicting the optimal noise level for SR. The resonance peak observed from the three metrics closely matches both the experimental measurements and the theoretical prediction based on Kramers time. Finally, we show a proof-of-concept demonstration of leveraging SR in ferroelectric PZT capacitors to enable signal recovery in FSK modulated communication. 




\section*{Methods}
\subsection*{Material}
Lead zirconium titanate (PZT) capacitor samples (purchased from Radiant Technologies) are used for the study of stochastic resonance (SR). The PZT device has the following layer structure: SiO\textsubscript{2} (500~nm) / TiO\textsubscript{2} (40~nm) / Pt (150~nm) / PZT (255~nm) / Pt (100~nm), with different top electrode area.

\subsection*{Electrical measurements}
The electrical measurements are performed in a probe station using a Multi System Analyzer 500 equipped with a Precision Multiferroic II Ferroelectric Tester. Prior to each measurement—i.e., feeding a sub-coercive signal combined with noise to the PZT device—the sample is poled by applying two opposite poling pulses with amplitude greater than $V_c$ to ensure the same reference remnant polarization ($P_r$) and coercive voltage ($V_c$) at the start. More detailed measurement protocols are provided in the Supplementary Information.

\subsection*{Data analysis}
The numerical simulations of Landau theory for our PZT device and the output polarization recordings of experiment and simulation are analyzed using in-house developed code in MATLAB and Python. The functions cov, bandpower, and pwelch in MATLAB are used to estimate cross-covariance, output power, and snr respectively.


\section*{Acknowledgments}
P.N. acknowledges the Start-up Grant from IISc, the Infosys Young Researcher Award, and funding from SERB (DST), New Delhi, Government of India (CRG/2022/003506), as well as support from the DST-COE on piezoMEMS (DST/TDT/AM/2022/084). All the authors acknowledge the use of the National Nanofabrication Centre, the Micro and Nano Characterization Facility, and the Advanced Facility for Microscopy and Microanalysis at IISc for various fabrication and characterization studies. A.A  acknowledges funding from SERB (DST), New Delhi, Government of India (MTR/2021/000823, CRG/2022/008128). A.A.,  M.R.S., K.R.J., and J.N. acknowledge the use of the Central Instrumentation Facility and the High Performance Computing facilities at IIT Palakkad.

\section*{Author Contributions}

V.D., T.B., P.N., and A.A. conceptualized the project. V.D., P.N., and T.B. designed the experiments, while V.D., T.B. and J.N. performed the measurements and data analysis related to the stochastic resonance experiments. V.D., T.B. and M.R.S developed the theoretical simulation framework. V.D. set up the FSK experiment and preformed data analysis.   M.R.S., K.R.J., J.N. assisted in device electrical measurements and simulation.  M.R.S., K.R.J., and J.N. also performed the experiments and data analysis related to Kramers time. A.A. contributed to the theoretical framework and data interpretation. All authors discussed the results and contributed to the final manuscript.

Correspondence should be addressed to P.N. (pnukala@iisc.ac.in) and A.A. (arvindajoy@iitpkd.ac.in).

\newpage

\newpage
\begin{center}
    \LARGE \textbf{Supplementary Information} \\[2em]
    
    \LARGE \textbf{Synchronous polarization switching at sub-coercive fields through stochastic resonance in ferroelectric thin-film capacitors} \\[2em]
    
    \large
    Vivek Dey$^{1\#}$, Thejas Basavarajappa$^{2\#}$, Manikantan R.S.$^3$, Kevin Renji Jacob$^{1,3}$,\\
    Jonnalagadda Nikhila$^{1,3}$, Arvind Ajoy$^3$, Pavan Nukala$^1$ \\[1em]
    
    \normalsize
    $^1$Centre for Nanoscience and Engineering, Indian Institute of Science, 560012, India \\
    $^2$School of Electrical and Computer Engineering, Cornell University, Ithaca, NY 14853, USA\\
    $^3$Department of Electrical Engineering, Indian Institute of Technology Palakkad, 678623, India \\
    [0.5em]
    $^\#$These authors contributed equally to this work \\[0.5em]
    \textit{Corresponding authors: pnukala@iisc.ac.in, arvindajoy@iitpkd.ac.in}
\end{center}

\SupplementaryMaterials

\setcounter{equation}{0}
\section*{\underline{Supplementary Note 1}: Mathematical description of stochastic polarization switching dynamics in a double-well potential in the presence of noise using Time Dependent Ginzberg-Landau (TDGL) formulation}

The Landau free energy density of a ferroelectric system is given by
\begin{equation}
    F = \alpha P^2 + \beta P^4 - P E
\end{equation}

Since we want to model the polarization switching in presence of noise, the corresponding time-dependent Ginzburg–Landau (TDGL) equation in the presence of the Langevin force $\xi(t)$ (which reflects the thermodynamic fluctuations due to thermal noise or external driving noise) is:

\begin{equation}
    \rho \frac{\partial P}{\partial t} = -\frac{\partial F}{\partial P} + \xi(t)
\end{equation}

where $\rho$ is the internal resistivity (a dissipative term) and $\xi(t)$ is white Gaussian noise.

The noise $\xi(t)$ has the following properties:
\begin{enumerate}
    \item $\langle \xi(t) \rangle = 0$
    \item The autocorrelation is $\langle \xi(t) \xi(t + \delta t) \rangle = 0$ if $\delta t > \tau$, where $\tau$ is the maximum time over which the Gaussian noise has any correlation.
    \item The correlation decay rate is independent of time such that $\langle \xi(t) \xi(t + \delta t) \rangle$ depends only on $\delta t$.
\end{enumerate}

The behavior of the system depends more on the cumulative effect of the random forcing $\xi(t)$, i.e., the integral over some time period long compared to $\tau$, rather than its instantaneous value. We can break up the integral into segments each of duration $\tau$:

\begin{equation}
    \int_0^t \xi(t') \, dt' = \int_0^\tau \xi(t') \, dt' + \int_\tau^{2\tau} \xi(t') \, dt' + \int_{2\tau}^{3\tau} \xi(t') \, dt' + \cdots
\end{equation}

This integral is a sum of independent terms, each drawn from the same distribution. According to the Central Limit Theorem, the integral obeys a normal distribution with mean zero (since $\langle \xi(t) \rangle = 0$), and its standard deviation scales with $\sqrt{t}$.

The integral in (3) has the properties of a Wiener process. Hence, the stochastic process of polarization switching in presence of a Langevin force $\xi(t)$ can be related to an underlying Brownian or Wiener process $W(t)$ as:

\begin{equation}
    \xi(t) = B \frac{dW(t)}{dt}
\end{equation}

where $B$ is a constant. The Wiener process $W(t)$ has the following properties:
\begin{enumerate}
    \item $W(0) = 0$ with probability 1
    \item Independent increments: For all $r < s \le t < u$, $W(u)-W(t)$ is independent of $W(s)-W(r)$
    \item $W(t)-W(s) \sim \mathcal{N}(0, t-s)$
    \item $W(t)$ is continuous in $t$
\end{enumerate}

According to the fluctuation-dissipation theorem, we can write:

\begin{equation}
    \langle \xi(t) \xi(t') \rangle = \frac{2k_B T}{V} \rho \delta(t - t') = \frac{2k_B T}{t_F A_F} \rho \delta(t - t')
\end{equation}

where $t_F$ is the thickness and $A_F$ is the area of the ferroelectric (PZT) capacitor.

Also,
\begin{equation}
    \langle \xi(t) \xi(t') \rangle = B^2 \left\langle \frac{dW(t)}{dt} \frac{dW(t')}{dt'} \right\rangle = \frac{2k_B T}{t_F A_F} \rho \delta(t - t')
\end{equation}

which implies,
\begin{equation}
    B = \sqrt{ \frac{2k_B T \rho}{t_F A_F} }
\end{equation}

Substituting in equation (4) gives,
\begin{equation}
    \xi(t) = \sqrt{ \frac{2k_B T \rho}{t_F A_F} } \frac{dW(t)}{dt}
\end{equation}

In our problem, the stochastic variable is the polarization $P$. We describe the system using a stochastic framework that captures the evolution of the macroscopic variable $P$ and its random fluctuations. To describe the evolution of the probability distribution $w(P,t)$ of the polarization fluctuations, we derive the Fokker–Planck equation.

From Fick's law of diffusion, the diffusion current is:
\begin{equation}
    J_{\text{diff}} = -D \frac{\partial w(P,t)}{\partial P}
\end{equation}

where $D$ is the diffusion coefficient.

The drift current is given by:
\begin{equation}
    J_{\text{drift}} = v w(P,t)
\end{equation}

where $v = \frac{\partial P}{\partial t} = -\frac{1}{\rho} \frac{\partial F}{\partial P}$ is the velocity of a representative point $P$ along the polarization axis in absence of thermal fluctuations.

From the continuity equation:
\begin{align}
    \frac{\partial w(P,t)}{\partial t} &= -\frac{\partial}{\partial P} \left( J_{\text{diff}} + J_{\text{drift}} \right) \nonumber \\
    &= -\frac{\partial}{\partial P} \left( v w(P,t) - D \frac{\partial w(P,t)}{\partial P} \right) \nonumber \\
    &= \frac{1}{\rho} \frac{\partial}{\partial P} \left[ \frac{\partial F}{\partial P} w(P,t) + \rho D \frac{\partial w(P,t)}{\partial P} \right] \nonumber \\
    &= \frac{1}{\rho} \frac{\partial}{\partial P} \left[ \frac{\partial F}{\partial P} w(P,t) + D' \frac{\partial w(P,t)}{\partial P} \right]
\end{align}

In statistical equilibrium, $\frac{\partial w(P,t)}{\partial t} = 0$, and $w(P,t)$ must reduce to a Boltzmann distribution:
\[
w(P,t) = w_0 \exp\left( -\frac{V F}{k_B T} \right)
\]
where $V$ is the volume, $F$ is the free energy density, $k_B$ is the Boltzmann constant, and $T$ is the temperature.

Substituting this into Eq.~(10), we obtain:
\[
D = \frac{k_B T}{V} = \frac{k_B T}{t_F A_F}
\]

Now introducing an external noise voltage (rms value $V_{\text{noise rms}}$) with flat power spectral density over a bandwidth $\Delta f_{BW}$, we modify Eq.~(7) as:

\begin{equation}
    \xi(t) = \sqrt{2\rho D_{\text{int}}} \frac{dW(t)}{dt} + \sqrt{2\rho D_{\text{ext}}} \frac{dW(t)}{dt}
\end{equation}

where
\[
D_{\text{ext}} = \dfrac{V_{noise\; rms}^2}{2\rho t_F^2 \Delta f_{BW}}
\]

In our experiments and simulation, we assume $D_{\text{int}} \ll D_{\text{ext}}$ and solve the discretized form of Eq.~(2) using the Euler–Maruyama method:

\begin{equation}
    P[i] = P[i-1] - \frac{\Delta t}{\rho} \left. \frac{\partial F}{\partial P} \right|_{i-1} + \sqrt{ \frac{2 D_{\text{ext}}}{\rho} } \Delta W[i-1]
\end{equation}

\section*{\underline{Supplementary Note 2}: Gaussian white noise generation and experimental measurement details}

We first generate Gaussian white noise of zero mean and unit variance. Scaling the noise by a factor of $n$ (i.e., $n \times \text{noise}$) results in a linear scaling of the standard deviation by the same factor $n$. Based on this property, we generate arrays of normally distributed random numbers using the \texttt{randn} function in \textsc{MATLAB}. These arrays are subsequently scaled by appropriate factors to obtain white noise signals with the desired standard deviation.

In this work, we use standard deviation values ranging from 0.15 to 5.10, with a step size of 0.15  for sample A and 0 to 1.9, with a step of 0.1 for sample B (more details of the parameters are listed in Table ST1). Fig. S2 illustrates the procedure for Gaussian white noise generation in \textsc{MATLAB}. The synthesized signal (input + noise) is applied to the PZT device using a Radiant ferroelectric test system.

\section*{\underline{Supplementary Note 3}: Poling protocols}
Ferroelectric samples exhibit an imprint effect, which can lead to an internal bias and asymmetry in the polarization switching behavior. To mitigate this and ensure consistent initial conditions, a poling voltage greater than twice the coercive voltage (\(2V_c\)) is applied prior to each measurement. The poling pulse, along with the sub-coercive signal and noise, is shown in Fig. S2.

\section*{\underline{Supplementary Table ST1}: Experimental parameters}
\begin{table}[H]
\centering
{%
\begin{tabular}{lcc}
\hline
\textbf{Parameter} & \textbf{Sample A} & \textbf{Sample B} \\
\hline
Signal Voltage, $V_{sub}$ ($V$)  & $1.575$ & $1$\\
Signal Frequency, $f_{signal}$ ($Hz$) & $75, 100, 150$ & $100$\\
No. of Cycles, $n$ & $75$ & $33$\\
Standard Deviation, $sd$ ($V$) & $[0, 5.1]$ step $0.15$ & $[0, 1.9]$ step $0.1$\\
Sampling Frequency, $f_s$ ($Hz$)& $10^4$ & $10^5$\\
Bandwidth, $f_{BW}$ ($Hz$)& $10^4$ & $10^5$\\
Ensemble, $ens$ & $1$ & $10$\\
\hline
\end{tabular}%
}
\caption*{\textbf{Table ST1:} Experimental parameters used for the measurements in Sample A and Sample B.}
\end{table}

\section*{\underline{Supplementary figures}:}
\begin{figure*}[!h]
\centering
\includegraphics[width=\textwidth]{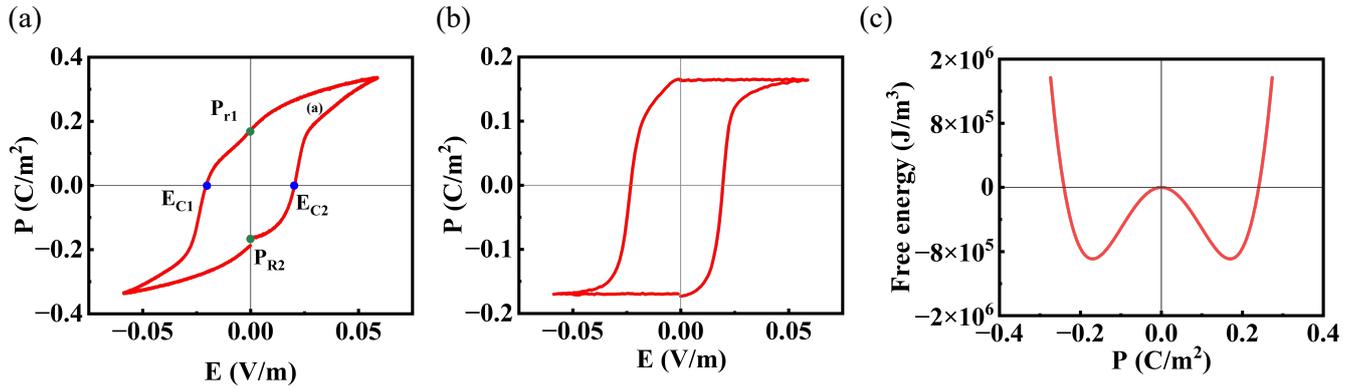}
\caption{\textbf{PE hysteresis and derived potential well for Sample-A}:(a) Polarization vs. electric field hysteresis loop measured for sample A. (b) PUND (Positive UP and NEgative Dowm) hysteresis loop. (c) Free energy vs. polarization plot representing the bistable potential well of the PZT device, modeled using Landau theory at zero bias.}
\label{fig:PEsampleA}
\end{figure*}

\begin{figure*}[!h]
\centering
\includegraphics[width=\textwidth]{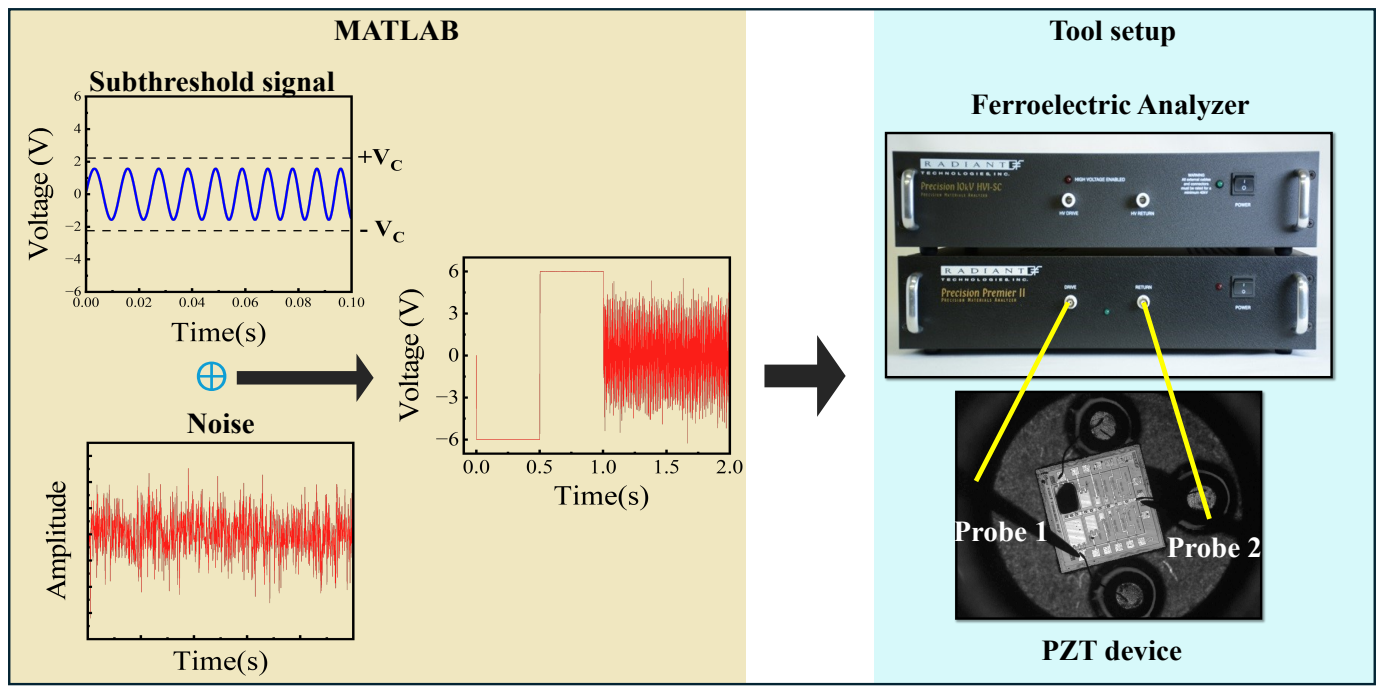}
\caption{\textbf{Noise generation and experimental procedure for ferroelectric SR measurement}:The input sub-coercive signal and the white gaussian noise is generated in MATLAB and these are added. An initial poling pulse is added to the start of every measurement to pole the PZT ferroelectric capacitor and switch to one of the two wells.}
\label{fig:Noise_gen}
\end{figure*}

\begin{figure*}[!h]
\centering
\includegraphics[scale=0.9]{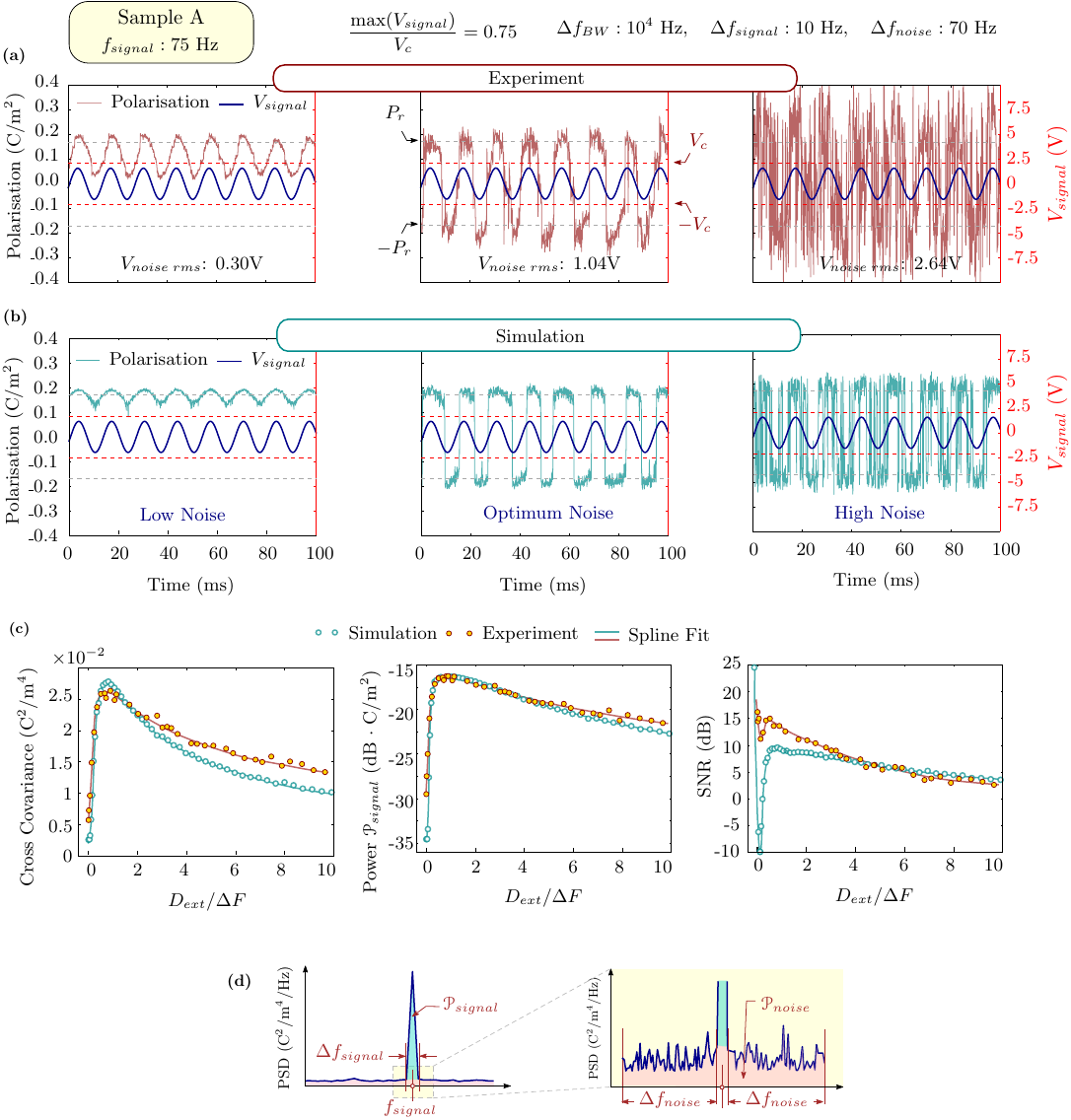}
\caption{\textbf{Polarization response and three metrics for Sample-A at $f_{signal}$ = 75 Hz}: (a) Experiment and (b) Simulated sub-coercive field polarization switching response at different $V_{noise\; rms}$ values at $f_{signal} = 75$ Hz. (c) Three independent FOMs showing accurate match between experiment and modeling. (d) Extraction method of background noise and signal power for signal-to-noise ratio estimation. }
\label{fig:SampleA75Hz}
\end{figure*}

\begin{figure*}[!b]
\centering
\includegraphics[scale=0.9]{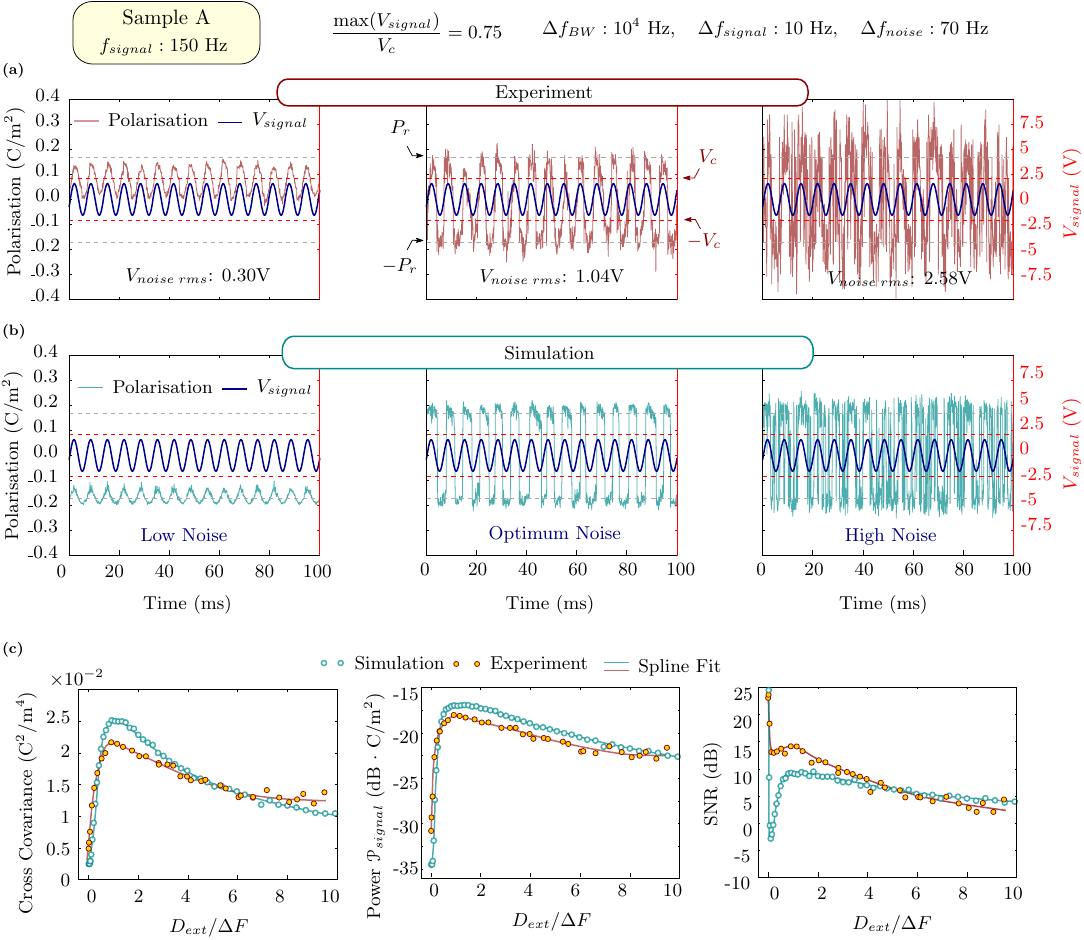}
\caption{\textbf{Polarization response and three metrics for Sample-A at 150 Hz}: (a) Experiment and (b) Simulated sub-coercive field polarization switching response at different $V_{noise\; rms}$ values at $f_{signal} = 150$ Hz. (c) Three independent FOMs showing accurate match between experiment and modeling. }
\label{fig:SampleA150Hz}
\end{figure*}

\begin{figure*}[!b]
\centering
\includegraphics[scale=0.9]{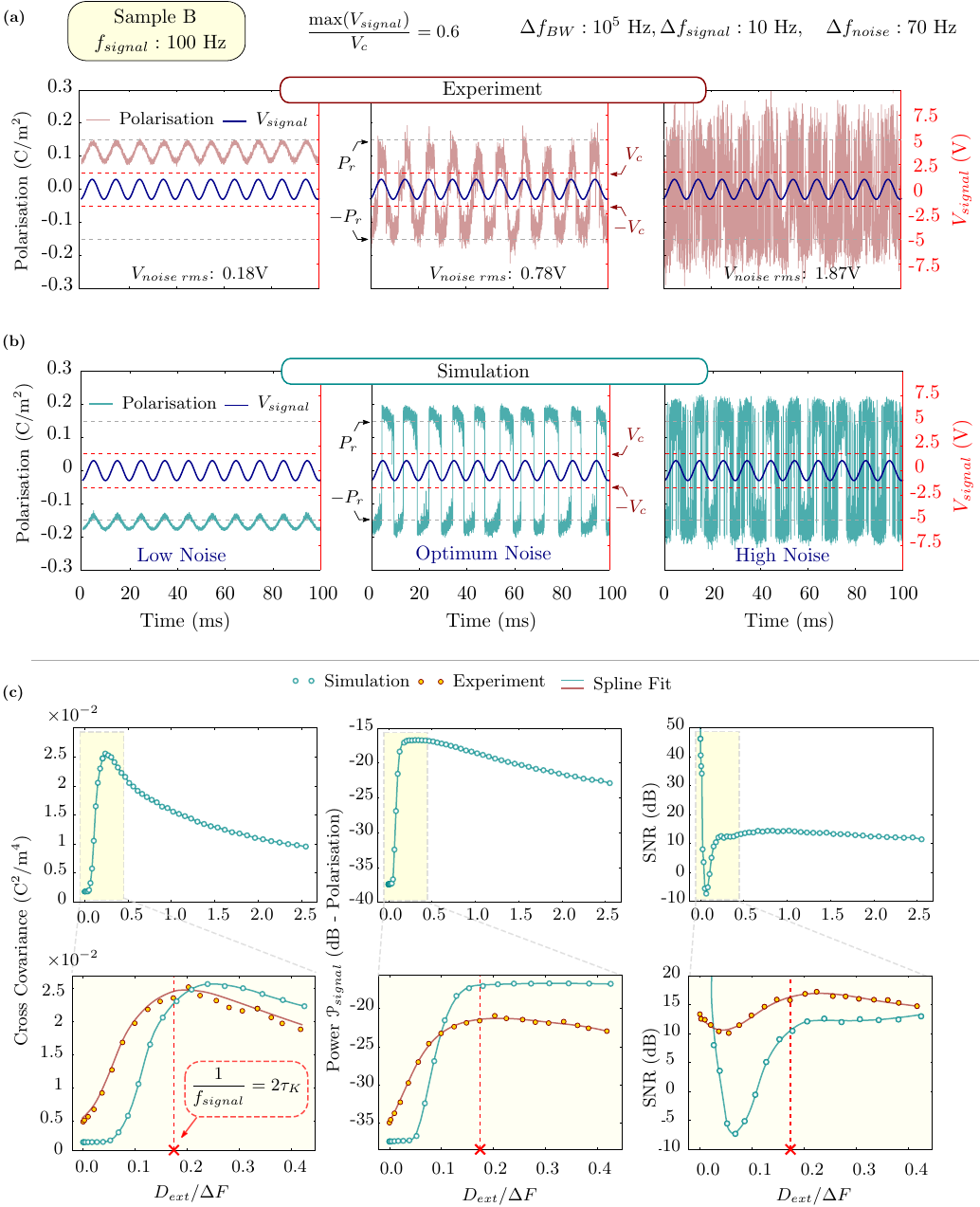}
\caption{\textbf{Polarization response and three metrics for Sample-B at 100 Hz.}(a) Simulated and (b) Experimental sub-coercive field polarization switching response at different $V_{noise\; rms}$ values at $f_{signal} = 100$ Hz. (c) Three independent FOMs showing accurate match between experiment and modeling. The second panel of (c) shows the zoomed-in view of the yellow region highlighted in the first-panel figures.}
\label{fig:SRsampleB}
\end{figure*}


\end{document}